\documentclass[12pt,nofootinbib]{article}
\usepackage{amssymb}
\usepackage{graphicx}
\usepackage{graphics}
\usepackage{amsmath}
\usepackage{latexsym}

\newcommand{\be}{\begin{equation}}
\newcommand{\ee}{\end{equation}}
\newcommand{\ba}{\begin{align}}
\newcommand{\ea}{\end{align}}
\newcommand{\bg}{\begin{gather}}
\newcommand{\eg}{\end{gather}}
\newcommand{\bseq}{\begin{subequations}}
\newcommand{\eseq}{\end{subequations}}

\makeatletter
\def\gsim{\compoundrel>\over\sim}
\def\lsim{\compoundrel<\over\sim}
\def\compoundrel#1\over#2{\mathpalette\compoundreL{{#1}\over{#2}}}
\def\compoundreL#1#2{\compoundREL#1#2}
\def\compoundREL#1#2\over#3{\mathrel
         {\vcenter{\hbox{$\m@th\buildrel{#1#2}\over{#1#3}$}}}}
\makeatother

\begin{document}

\title{On sgoldstino interpretation of the diphoton excess}

\author{S.~V.~Demidov$^{1,}$\thanks{{\bf e-mail}: demidov@ms2.inr.ac.ru}\,,
D.~S.~Gorbunov$^{1,2,}$\thanks{{\bf e-mail}: gorby@ms2.inr.ac.ru}\,,
\\
$^{1}${\small{\em
Institute for Nuclear Research of the Russian Academy of Sciences,
}}\\
{\small{\em
60th October Anniversary prospect 7a, Moscow 117312, Russia
}}\\
$^{2}${\small{\em
Moscow Institute of Physics and Technology,
}}\\
{\small{\em
Institutsky per. 9, 
  Dolgoprudny 141700, Russia
}}\\
}
\date{}

\maketitle

\vspace{-10cm}
\begin{flushright}
INR-TH-2015-036
\end{flushright}
\vspace{8.5cm}

\begin{abstract}
We point out that the diphoton excess at about 750~GeV recently
discovered by the LHC experiments can be explained within
supersymmetric models with low scale supersymmetry breaking with
sgoldstino as a natural candidate. We discuss phenomenological
consequences of this scenario describing possible signatures to test
this hypothesis.

\end{abstract}

\section{Introduction}
The first results obtained by the ATLAS and CMS collaborations in
proton-proton collisions at $\sqrt{s}=13$~TeV at LHC Run-II have been
recently revealed~\cite{seminar}. Among them there has been 
highlighted a small excess in searches for diphoton
resonances~\cite{ATLAS,CMS}. Although its local significance is not
very high, 3.9$\sigma$ for ATLAS and $2.6\sigma$ for CMS, it is most
exciting that both experiments observed the excess at the same
diphoton invariant mass around 750~GeV, and it is tempting to interpret
it as a signal from long-awaited new physics. Several studies have
already been performed in this direction~\cite{others} and among them
are explanations within a low scale supersymmetry
framework~\cite{Bellazzini:2015nxw,Petersson:2015mkr}. In this class of 
models (see,
e.g.~\cite{Brignole:2003cm,Petersson:2011in,Antoniadis:2012ck,Dudas:2012fa}) 
apart from usual superpartners of the Standard Model particles, the low 
energy theory contains also a part of the sector responsible for
supersymmetry breaking. In a minimal scenario, this part contains
Goldstone fermion -- goldstino -- and its superpartners dubbed
sgoldstinos. Interactions of the latter with the Standard Model
particles are determined by soft supersymmetry breaking parameters and 
suppressed by the supersymmetry breaking scale $\sqrt{F}$. Collider
phenomenology of sgoldstinos with masses of order electroweak scale
have been discussed for instance
in~\cite{Dudas:2012fa,Perazzi:2000ty,Gorbunov:2002er,Demidov:2004qt,Gorbunov:2000ht,Bellazzini:2012mh,Petersson:2012nv,Astapov:2014mea,Petersson:2015rza}. In
Refs.~\cite{Bellazzini:2015nxw,Petersson:2015mkr} it has been proposed
that sgoldstino (scalar, pseudoscalar or both) is responsible for the
diphoton excess at $750$~GeV. In this note we further discuss
phenomenological implications of this proposal pointing out at
particular signatures which can be used to verify this scenario.


\section{Exploring the model}
The scalar $S$ sgoldstino interactions with photons and gluons are
governed by the following effective lagrangian (see,
e.g. Ref.~\cite{Gorbunov:2001pd})  
\be
\label{eq:1}
{\cal L}_1 =  - \frac{M_3}{2\sqrt{2}F} \, S \, G^a_{\mu \nu}
G^{a\mu \nu} - \frac{M_{\gamma \gamma}}{2\sqrt{2}F}  \, S \, F_{\mu \nu}
F^{\mu \nu}\,,
\ee
where 
$M_{\gamma\gamma} = M_1 \, \cos^2{\theta_W} + M_2\,\sin^2{\theta_W}$ 
and $M_{1,2,3}$ are soft gaugino masses of a supersymmetric extension
of the Standard Model. 
The dominating production mechanism of sgoldstino at high
energy $pp$ collisions is gluon-gluon
fusion~\cite{Perazzi:2000ty,Gorbunov:2002er} for typical hierarchy of
soft supersymmetry breaking parameters: gluino is the heaviest gaugino
while the soft trilinear couplings are of the same size as $M_i$.  
The production cross section $\sigma_S$ is related to the decay of
sgoldstino 
to gluons $\Gamma(S\to gg) = \frac{M_3^2m_S^3}{4\pi F^2}$ which is
typically the main decay channel for $\sqrt{F}$ at TeV scale and $m_S < 
\sqrt{F}$ which we will assume in what follows. The width of 
sgoldstino decay into photons is given by $\Gamma(S\to\gamma\gamma) =
\frac{M_{\gamma\gamma}^2m_S^3}{32\pi F^2}$.
Apart from $gg$ and $\gamma\gamma$ the most important decay modes 
of sgoldstino relevant for our study are into $WW$, $ZZ$ and
$Z\gamma$. Corresponding interactions are 
\be
\label{eq:2}
{\cal L}_2 = -\frac{M_2}{\sqrt{2}F}SW^{\mu\nu}W_{\mu\nu} - 
\frac{M_{ZZ}}{2\sqrt{2}F}SZ^{\mu\nu}Z_{\mu\nu} 
- \frac{M_{Z\gamma}}{\sqrt{2}F}SZ^{\mu\nu}F_{\mu\nu} 
\ee
with $M_{ZZ} = M_1 \, \sin^2{\theta_W} + M_2\,\cos^2{\theta_W}$ and 
$M_{Z\gamma} = (M_2-M_1)\sin{\theta_W}\cos{\theta_W}$.

To explain the diphoton excess we require that the mass of sgoldstino
is equal to 750~GeV and
\be 
\label{excess}
3~{\rm fb}\lsim\sigma_{\gamma\gamma}\lsim 13~{\rm fb}\;\;
{\rm at}\;\; 13~{\rm TeV}, 
\ee 
where we define
\be
\label{def}
\sigma_{\gamma\gamma} \equiv \sigma_{S}\times {\rm Br}(S\to\gamma\gamma)
\ee
and as $gg$ decay mode is dominant,
\be
{\rm Br}(S\to\gamma\gamma) =
\frac{\Gamma(S\to\gamma\gamma)}{\Gamma(S\to gg)}.
\ee
The information about the width of this resonance is still quite
uncertain and we will not be focusing on it in the present study.
One should check that this scenario is phenomenologically viable and
passes all constraints obtained in previous collider searches.
Let us discuss the compatibility of new diphoton excess with
the ATLAS and CMS results at $\sqrt{s}=8$~TeV. The strongest bound
$\sigma_{\gamma\gamma} \lsim 1.5$~fb at this collision energy was
obtained by the 
CMS~\cite{Khachatryan:2015qba} while somewhat weaker constraints come
from ATLAS data~\cite{Aad:2015mna}. The ratio of the production cross
sections for sgoldstino case of $\sqrt{s}=8$ and 13~TeV is almost
independent on parameters of the model and is found to be
$\frac{\sigma_S(\sqrt{s} = {13~{\rm  TeV}})}{\sigma_S(\sqrt{s} = 8~{\rm
    TeV})} \approx 5$.  
The excess~\eqref{excess} requires $0.6~{\rm
  fb}\lsim\sigma_{\gamma\gamma}(\sqrt{s}=13~{\rm TeV})\lsim 2.6$~fb
which seems to be still allowed partly by the present bounds. 
The dominating decay into gluon pairs reveals itself in a dijet
signature. The strongest present bound from the searches for dijet
resonances is $\sigma_S\times {\rm Br}(S\to gg) \lsim
30$~pb~\cite{jjboundAad:2014aqa}. 
Searches for resonances decaying
into pair of massive vector bosons $WW$, $ZZ$ have been performed
at the LHC Run-I~\cite{ZZboundAad:2015kna,WWboundAad:2015agg} and the 
strongest limits look 
\be
\sigma_{WW}\lsim 0.03~{\rm pb},\;\;\;
\sigma_{ZZ}\lsim 0.012~{\rm pb}\;\;\;\;
{\rm at}\;\;8~{\rm TeV}
\ee
with the definitions of $\sigma_{WW}$ and $\sigma_{ZZ}$ similar
to~\eqref{def}. It is 
important to note that 
the couplings of sgoldstino to $Z$ and 
$W$ bosons are not independent from that of to photons, see
Eqs.~\eqref{eq:1} and~\eqref{eq:2}. Thus, in
general one expects that sgoldstino interpretation of the diphoton 
resonance will result in certain predictions with the diboson ($WW$,
$ZZ$ or $Z\gamma$) signatures. We explore this possibility by performing a
scan  over relevant parameter space. Namely we fix the supersymmetry
breaking scale $\sqrt{F}=5$ or 7~TeV while take gaugino masses
randomly in the following ranges: 0.2~TeV$<M_{1,2}<\sqrt{F}$ and
1.7~TeV$<M_{3}<\sqrt{F}$. The upper bounds in these intervals come
from perturbative unitarity constraints of the effective sgoldstino
interactions while the lower 
bounds are inspired by results of the direct searches\footnote{
Here we conservatively assume strong bound on gluino mass, see the
latest results~\cite{seminar}. Note, however, that it can be somewhat
relaxed for considered class of models with gravitino as the lightest
supersymmetric particle  when gluino undergoes multistage decays, see
e.g.~\cite{Aad:2015iea}.
} for gauginos. For each model characterized by chosen set of
parameters $M_{1,2,3}, F$ we calculate sgoldstino production  
cross section and its relevant decay widths using formulas of
Refs.~\cite{Spira:1997dg} and~\cite{Gorbunov:2002er}.
Then, we select phenomenologically accepted models which predict
diphoton rate in the interval~\eqref{excess} and also 
satisfy the experimental constraints from searches for resonances
in dijets, diphotons and double massive vector bosons we described
above. 

\begin{figure}[!htb]
\begin{center}
{\includegraphics[angle=-90,width=0.50\textwidth]{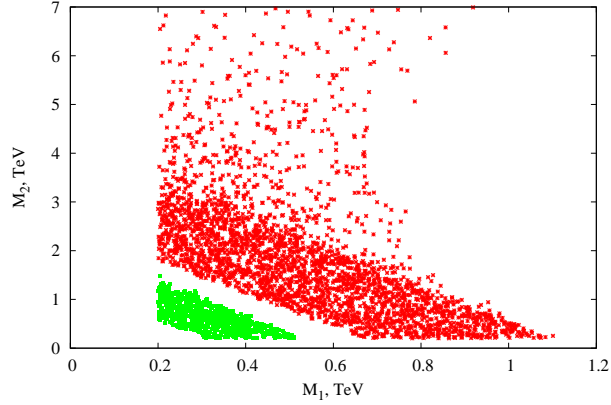}}
\end{center}
\caption{\label{M1_M2}  Scatter plot in $(M_{1},M_{2})$
 plane for $\sqrt{F}=5$~TeV (green/light gray) and $\sqrt{F}=7$~TeV
 (red / dark gray). 
}
\end{figure}

\begin{figure}[!htb]
\begin{center}
{\includegraphics[angle=-90,width=0.50\textwidth]{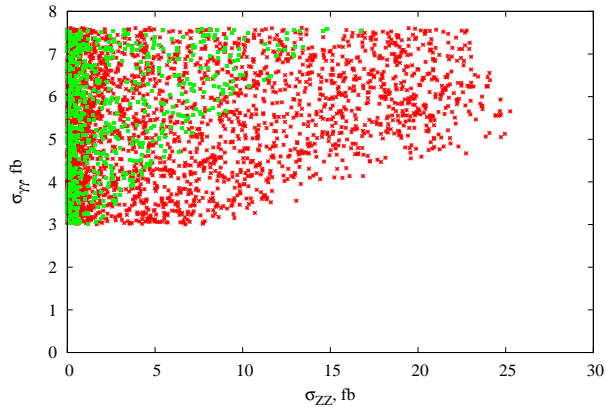}}
\end{center}
\caption{\label{ZZ_gg} Diphoton cross section versus $ZZ$ cross
  section for $\sqrt{F}=5$~TeV (green) and $\sqrt{F}=7$~TeV
  (red); we set $\sqrt{s}=13$~TeV. 
}
\end{figure}

\begin{figure}[!htb]
\begin{center}
{\includegraphics[angle=-90,width=0.50\textwidth]{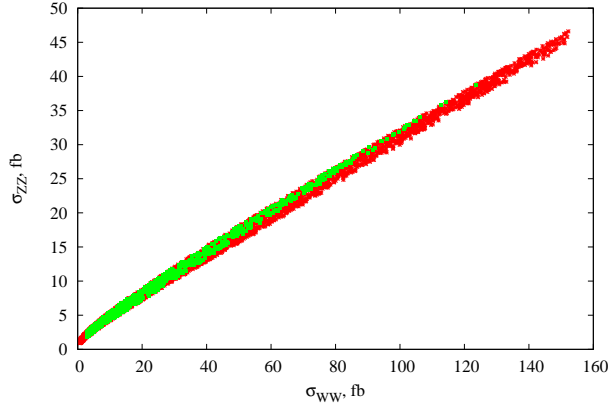}}
\end{center}
\caption{\label{ZZ_WW}  $ZZ$ and $WW$ cross sections for
  $\sqrt{F}=5$~TeV (green) and $\sqrt{F}=7$~TeV (red); we set
  $\sqrt{s}=13$~TeV.  
}
\end{figure}

\begin{figure}[!htb]
\begin{center}
{\includegraphics[angle=-90,width=0.50\textwidth]{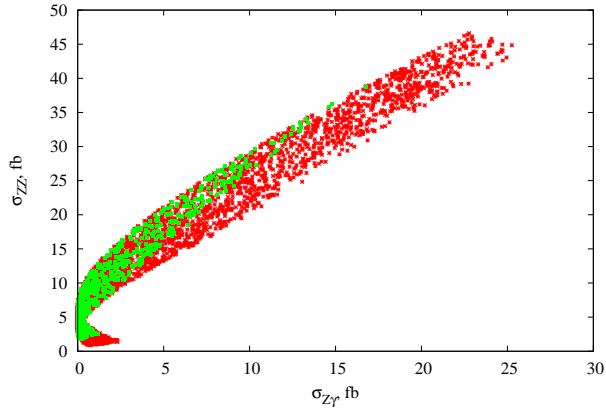}}
\end{center}
\caption{\label{ZZ_Zg}  $ZZ$ and $Z\gamma$ cross sections for
  $\sqrt{F}=5$~TeV (green) and $\sqrt{F}=7$~TeV (red); we set
  $\sqrt{s}=13$~TeV.  
}
\end{figure}

The results of the scan are presented in
Figs.~\ref{M1_M2}--~\ref{ZZ_Zg}. 
In Fig.~\ref{M1_M2} we show the parameters $M_{1,2}$ of the selected
models for different values of supersymmetry breaking scale
$\sqrt{F}$. As expected, larger $\sqrt{F}$ require an increase of
gaugino masses $M_{1,2}$ because sgoldstino couplings behave as
$M_i/F$. Similar correlation is found for $M_3$. Diphoton and $pp\to
S\to ZZ$ cross sections calculated at different $\sqrt{F}$ are
shown in Fig.~\ref{ZZ_gg}. 
We see that the bound on diphoton cross section obtained by CMS in the 
Run-I effectively cuts all the models with $\sigma_{\gamma\gamma}
\gsim 7.6$~fb. At the same time predicted values for the resonance
cross section with $ZZ$ final state $\sigma_{ZZ}$ reach values about
17~fb and 26~fb 
for $\sqrt{F}=5$~TeV and~7~TeV, respectively. In Figs.~\ref{ZZ_WW}
and~\ref{ZZ_Zg} we show obtained values of the resonance production
cross sections with  $ZZ$, $WW$  and $Z\gamma$ final states. 
The cross sections for these channels turn out to be of the
sizes reachable at the LHC, Run-II with larger statistics. Moreover
there are correlations between values of these cross sections which
appear  because of relations between corresponding sgoldstino coupling 
constants. Thus, the model with sgoldstino can be tested by  measuring 
different diboson cross sections which seems to be within the reach of
the 
Run-II experiments at least for their larger values. At the same time
lesser values of the cross sections $\sigma_{ZZ}$, $\sigma_{WW}$ and
$\sigma_{Z\gamma}$ correspond to small values of $M_{i}$. Thus, this
part of parameter space can be probed by direct searches for light
gauginos. 

Further, let us mention that we consider here only the case of scalar 
sgoldstino. Its pseudoscalar partner is also a viable candidate for
explanation of the diphoton excess, see 
also~\cite{Bellazzini:2015nxw,Petersson:2015mkr}. We can add that if
there is a mixing between scalar and 
pseudoscalar sgoldstinos this will result in changes of angular
distribution of decay products, photons or massive vector bosons. 
Scalar and pseudoscalar sgoldstino couplings to the SM particles and
hence the corresponding production cross sections
($\sigma_{\gamma\gamma}$, $\sigma_{ZZ}$ etc.) are closely related to
each other (see, e.g.~\cite{Gorbunov:2002er}). So, if the mixing is
reasonably small and their masses are different one expects to observe
$\gamma\gamma$--peak at another invariant mass associated with the
sgoldstino twin. Similar resonances are expected for other diboson
final states. 

Here we concentrated mainly on sgoldstino physics related to 
vector bosons. However, existence of the 750~GeV sgoldstino resonance
can have interesting implications related to the SM fermions. In
particular, corresponding sgoldstino couplings are determined by soft
trilinear coupling constants $A_{ij}^{U,D,L}$ which can have nontrivial flavor
structure. Thus, we can expect flavor violating processes mediated by
virtual sgoldstino: top-quark or heavy meson decays. Moreover, there
can be peaks at 750~GeV invariant mass of fermion-antifermion pair
(quarks or leptons) including those of different flavour.

We do not discuss here width of the sgoldstino resonance
$\Gamma_S$. For considered set of parameters of sgoldstino model its
value is always smaller than the GeV scale. Large width $\Gamma_S\sim 
46$~GeV which is somewhat favorable by the ATLAS 
results~\cite{ATLAS} (but not the CMS) seems not to be allowed within
the considered framework. However, in general the width of sgoldstino
can be increased to some extent in the case of nonminimal
supersymmetric extensions involving light singlet scalar (e.g. in
NMSSM). In this case sgoldstino can have a considerable decay  rate
into pair of the scalars while the hierarchy between branchings of the
decays into vectors bosons will be almost unchanged.


\section{Discussion and conclusions}
To summarize, we argue that sgoldstino with the mass about 750~GeV is
a natural candidate for explanation of the small diphoton excess
observed recently by the ATLAS and CMS experiments. Typical values of 
supersymmetry breaking scale and soft gaugino masses required for this
explanation are found to be below 10~TeV. We explore possible range of
the sgoldstino production cross section with the diboson final states
$WW$, $ZZ$ and $Z\gamma$ and found that these processes as well as
direct searches for gauginos can be used to test the hypothesis about 
750~GeV sgoldstino. 

Finally, let us note that in the class of models with low scale
supersymmetry breaking the lightest supersymmetric particle is
gravitino.  For values of $\sqrt{F}$ considered in this note its mass
is about $m_{3/2} = \sqrt{\frac{8\pi}{3}}\frac{F}{M_{Pl}} \sim
10^{-3}-10^{-2}$~eV. This region is allowed by astrophysical
bounds from Supernovae~\cite{Grifols:1996wq} and by cosmological
constraints from Big Bang Nucleosynthesis~\cite{Moroi:1993mb}.

\section*{Acknowledgments}
{\footnotesize
The work is supported by the RSCF grant 14-12-01430.
}


\end{document}